\documentclass[aps,prd,twocolumn,showpacs,preprintnumbers,
superscriptaddress,nofootinbib]{revtex4}

\usepackage{slashed}

\usepackage{latexsym}
\usepackage{bm}

\usepackage{graphicx}
\usepackage{amssymb,amsmath}

\newcommand\beq{\begin{equation}}
\newcommand\eeq{\end{equation}}

\newcommand\nubar{{\bar\nu}}
\newcommand{\Psibar}{{\bar \Psi}}
\newcommand{\photino}{{\tilde \gamma}}
\newcommand{\kbar}{{\bar k}}
\newcommand{\rarr}{\rightarrow}

\begin{document}


\preprint{}

\title{Electroweak Bremsstrahlung in Dark Matter Annihilation}

\author{Nicole F.\ Bell} 
\affiliation{School of Physics, The University of Melbourne, 
Victoria 3010, Australia}

\author{James B.\ Dent}
\affiliation{Department of Physics and Astronomy,
Vanderbilt University, Nashville, TN 37235, USA}

\author{Thomas D.\ Jacques}
\affiliation{School of Physics, The University of Melbourne, 
Victoria 3010, Australia}

\author{Thomas J.\ Weiler}
\affiliation{Department of Physics and Astronomy,
Vanderbilt University, Nashville, TN 37235, USA}

\date{\today}

\begin{abstract}
  A conservative upper bound on the total dark matter (DM)
  annihilation rate can be obtained by constraining the appearance
  rate of the annihilation products which are hardest to detect.  The
  production of neutrinos, via the process $\chi \chi \rightarrow
  \bar\nu \nu $, has thus been used to set a strong general bound on
  the dark matter annihilation rate.  However, Standard Model
  radiative corrections to this process will inevitably produce
  photons which may be easier to detect.  We present an explicit
  calculation of the branching ratios for the electroweak
  bremsstrahlung processes $\chi \chi \rightarrow \bar\nu \nu Z$ and
  $\chi \chi \rightarrow \bar\nu e W$.  These modes inevitably lead to
  electromagnetic showers and further constraints on the DM
  annihilation cross-section.  In addition to annihilation, our
  calculations are also applicable to the case of dark matter decay.
\end{abstract}



\maketitle


\section{Introduction}

\noindent
The identity of the dark matter (DM) is one of the great unresolved
questions in particle physics and
cosmology~\cite{Kamionkowski_review,Bertone_review,Bergstrom_review}.
An important method of probing DM properties is via indirect
detection, whereby we look for the appearance of particles produced
via DM annihilation or decay.  We can search for such a signal
emanating from the dark matter concentration in our own galaxy, other
galaxies (satellite, dwarf, or clustered), 
or for a isotropic flux from the dark matter distributed
throughout the Universe~\cite{Profumo}.
Investigated signals include positrons, gamma-rays, x-rays, and even
microwaves (the ``WMAP haze''~\cite{Fink}).

If we make the reasonable assumption that DM decay or annihilation
products must be Standard Model (SM) particles (i.e. we assume the
dark matter is the lightest stable particle in the beyond-SM sector)
then it is possible to set a conservative upper bound on the {\it
  total} DM annihilation rate by looking for the annihilation products
which are the hardest to detect, namely, neutrinos~\cite{BBM}.  All
other possible final states would lead to the production of gamma
rays, for which more stringent bounds apply.  For example, quarks and
gluons hadronize, producing pions and thus photons via
$\pi^0\rightarrow \gamma\gamma$; the decays of $\tau^\pm$, $W^\pm$,
and $Z^0$ also produce $\pi^0$.  Charged particles produce photons via
electromagnetic radiative corrections~\cite{RC,BergstromFolk},
while energy loss processes for $e^\pm$ also produce
photons~\cite{Eloss}.
By calculating the cosmic diffuse neutrino flux produced via the DM
annihilation process $\chi\chi\rightarrow \bar\nu \nu$ in all halos
throughout the Universe, a strong and general bound on the DM total
annihilation cross section has recently been derived~\cite{BBM}.  The
corresponding signal from our own galaxy can be used to set a
comparable limit (and improves upon the cosmic bound in some mass
ranges)~\cite{Yuksel} while the technique has been extended to low
(MeV) masses in Ref.~\cite{Sergio}.  Analogous bounds have been derived
for the DM decay rate~\cite{Sergio_decay}.

\begin{figure}[t]
\centering
\includegraphics[width=3.0in]{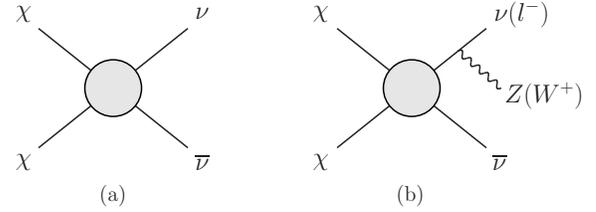}
\caption{
  The lowest order tree level process $\chi\chi\rightarrow \bar\nu
  \nu$ (left) is accompanied by electroweak bremsstrahlung processes
  (right).
\label{fig:basic}}
\end{figure}

The general upper bound on the {\it total} DM annihilation cross
section defined via the limit on $\chi\chi\rightarrow \bar\nu \nu$ is
surprisingly strong. (See Ref.~\cite{MJBBY} for a comparison between
photon-based and neutrino-based limits.)  However, a scenario in which
neutrinos alone are produced in the final state is technically
impossible.  Even leaving aside the theoretical issue that a direct coupling 
of DM to only neutrinos violates the $SU(2)$-invariance of the 
weak interaction, 
electroweak radiative corrections imply indirect couplings
to states other than neutrinos.  For example, for energies above $M_{W,Z}$,
electroweak bremsstrahlung of $W$ or $Z$ bosons can occur at sizeable
rates~\cite{Kachelriess,Berezinsky}, see Fig.~\ref{fig:basic}.  The
hadronic decays of these gauge bosons produce neutral pions, which
decay to gamma rays.  Even for energies below $M_W$, processes
involving virtual electroweak gauge bosons will lead to particles with
electromagnetic interactions, though the rate for such processes is
suppressed at low energy.

Kachelriess and Serpico have estimated the constraints on the cross
section for $\chi\chi\rightarrow \bar\nu \nu$ (and hence on total DM
annihilation cross section) by considering gamma rays produced via the
accompanying process $\chi\chi\rightarrow \bar\nu \nu
Z$~\cite{Kachelriess}.  We present here an explicit calculation of the
branching ratios for the electroweak bremsstrahlung processes
$\chi\chi\rightarrow \bar\nu \nu Z$ and $\chi\chi\rightarrow \bar\nu e
W$.

Note that the expected magnitude of the DM total annihilation cross
section varies enormously between specific models.  For LSP (lightest
supersymmetric particle) DM, s-wave annihilation to fermions is
helicity suppressed and thus the lowest order annihilation rate can be
quite small.  On the other hand, there is no such suppression for
Kaluza-Klein DM.  Please refer to the Appendix for a detailed
discussion of these issues.  Our analysis is not specific to any
particular dark matter candidate (nor do we assume the DM is a thermal
relic).

\section{W-strahlung}

\begin{figure}[t]
\centering
\includegraphics[width=0.8\columnwidth]{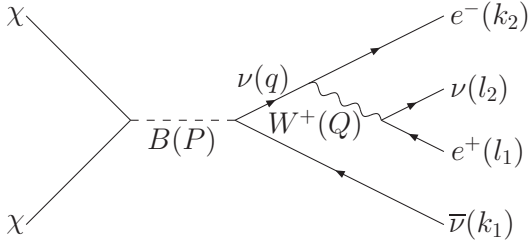}
\caption{
  Feynman diagram for the W-strahlung process $\chi\chi \rightarrow
  e^-\bar\nu W^{+*} \rightarrow e^+ e^- \bar\nu \nu$
\label{fig:wbrem}}
\end{figure}

We shall first consider DM annihilation to four body final states via
W-strahlung, an example of which is the $\chi\chi \rightarrow e^+ e^-
\bar\nu \nu$ process shown in Fig.~\ref{fig:wbrem}, and calculate the
ratio of the cross section for this process, to that for the lowest
order tree level process $ \chi \chi \rightarrow \bar\nu \nu$.  For
simplicity, we will assume that the coupling between the DM-current
and the neutrino-current is mediated by a scalar boson ``$B$''.  Given
this scalar coupling, the terms in the matrix element involving the
initial state ($\chi$) particles will factorize from the full matrix
element.  It is thus useful to consider the matrix element for the
decay of the virtual $B^*$.

We will first discuss the phase space calculation, and begin by noting
that it is useful to treat the process shown in Fig.~\ref{fig:wbrem}
as a sequence of three $1\rightarrow 2$ particle decays of a virtual
particle, viz., $B^*(P)\rightarrow \nu^*(q) + \bar{\nu}(k_1)$,
followed by $\nu^*(q)\rightarrow W^{*+}(Q) + e^-(k_2)$, followed by
$W^{*+}(Q)\rightarrow \nu(l_2) + e^+(l_1)$.  The four-body final state
Lorentz-Invariant Phase Space (LIPS) is given by
\begin{eqnarray}
LIPS^{(4)} &=& (2\pi)^4\int dk_2\int dk_1\int dl_2 \int dl_1\nonumber \\
&\times&\delta^4(P-l_1-l_2-k_1-k_2)\,,
\end{eqnarray}
where $dk\equiv (2\pi)^{-3}\,d^3 k/2k_0$, etc.
We integrate over the momenta of the virtual particles by inserting 
\begin{eqnarray}
d^4q\,d^4Q\,\delta^{4}(q-k_2-Q)\,\delta^4(Q-l_1-l_2),
\end{eqnarray}
which ensures momentum conservation for the virtual processes.  The
phase space then factorizes into a product of three separate two-body
phase space factors convolved over the two virtual particle momenta:
\begin{eqnarray}
LIPS^{(4)}=\int\frac{dq^2}{2\pi}\int\frac{dQ^2}{2\pi}\,
\int dLIPS^{(2)}(P^2,q^2,k_1^2)\nonumber \\
\times \int dLIPS^{(2)}(q^2,Q^2,k_2^2)\,\int dLIPS^{(2)}(Q^2,l_2^2,l_1^2)\,.
\end{eqnarray}
Each two-body differential phase space factor is easily evaluated in the 
respective two-body center of momentum (CoM) frame, using the expression
\begin{eqnarray}
dLIPS^{(2)}(x,y,z)= \frac{1}{8\pi}\frac{\sqrt{\lambda (x,y,z)}}{x}
   \left(\frac{d\overline{\Omega}}{4\pi}\right)\,,
\end{eqnarray}
where $d\overline{\Omega}$ is the CoM solid angle, 
and $\lambda(x,y,z)$ is the triangle function given by
\begin{eqnarray}
\lambda(x,y,z) = x^2 + y^2 + z^2 -2xy -2xz - 2yz\,.
\end{eqnarray}
The phase space may be written in a form useful for our calculation,
\begin{eqnarray}
LIPS^{(4)} 
&& \hspace{-6mm}
= \frac{1}{16} 
\frac{1}{(2 \pi)^4}
\int_{0}^{P^2}dq^2\int_{0}^{q^2}dQ^2
   \,\frac{(P^2-q^2)(q^2-Q^2)}{P^2\,q^2} \nonumber \\
   &\times& \left(\frac{d\overline{\Omega}_P}{4\pi}\right) 
   \,\left(\frac{d\overline{\Omega}_q}{4\pi}\right)
   \,\int dLIPS^{(2)}(Q^2,0,0)\,.
   \label{lips}
\end{eqnarray}
Here and throughout, we neglect the masses of the leptons, hence
$k_1^2 = k_2^2 = l_1^2 = l_2^2 = 0$.

We next calculate the matrix element for $B^*\rightarrow
\bar\nu_e\,e^- W^{+*}\rightarrow \bar\nu_e\,e^- \nu_e\,e^+$, which is
given by
\begin{eqnarray}
\label{matrixelement}
\mathcal{M}_W = g_B \frac{g^2}{2}\,
\left[\bar{u}(k_2)\gamma^{\mu}\frac{1-\gamma^5}{2}
\frac{\slashed{q}+m_{\nu}}{q^2 - m_{\nu}^2} v(k_1)\right] \nonumber\\
\times\left[\bar{u}(l_2)\gamma_{\mu}\frac{1-\gamma^5}{2}v(l_1)\right]
\left(\frac{-1}{Q^2-m_{W}^2+im_{W}\Gamma_W}\right)\,,
\end{eqnarray}
where the (non-standard) $B\nu\bar\nu$ and (standard) $W\nu e$ couplings are $g_B$ and
$g/\sqrt{2}=e/\sqrt{2}\,\sin\theta_w$, respectively.  We have expressed
the matrix element in Eq.~(\ref{matrixelement}) in Feynman t'Hooft
gauge, but note that our result is gauge invariant.
After squaring, summing over spins, and performing the integration
over the momenta $l_1$ and $l_2$, we obtain
\begin{eqnarray}\label{eq:1}
&\int&  dLIPS^{(2)}(Q^2,0,0) \sum_{spins}|\mathcal{M}_W|^2 = 
\Big[  8s\,(Q\cdot k_2)(Q\cdot q) \nonumber \\ 
& + & 4s\,Q^2\,(k_2\cdot q) -8q^2\,(Q\cdot k_2)(Q\cdot P) - 
4Q^2\,q^2\,(k_2\cdot P) \Big]  \nonumber \\
&\times& \frac{1}{3\cdot 2^6\,(2\pi)}\,
\frac{g_{B}^2\,g^4}{q^{4}\,[(Q^2-m_{W}^2)^2 + (m_W\Gamma_W)^2]}\,,
\end{eqnarray}
where $s\equiv P^2$ is the center of mass energy squared, and we have made use
of the identity
\beq
\int dLIPS^{(2)}(Q^2,0,0)\;l_1^\alpha l_2^\beta = 
\frac{1}{96\pi}\,[2 Q^\alpha Q^\beta +Q^2 g^{\alpha\beta}]\,.
\eeq
We now multiply this partial result by the remaining part of the phase
space in Eq.~(\ref{lips}), and perform the integrations over
$d\overline{\Omega}_P$, $d\overline{\Omega}_q$, and $q^2$, to obtain
the rate $\Gamma(\rightarrow \nu_e \bar{\nu}_e e^+ e^-)$.  We wish to
compare this rate with that for the $B^*\rightarrow\nu_e\bar\nu_e$,
for which the lowest-order tree-level expression is $\int dLIPS
\sum_{spins}|\mathcal{M}|^2 = g_{B}^2\,s/(4\pi)$.  The resulting
expression for the ratio of rates for these two processes is
\begin{eqnarray}
\label{hardsingle}
&&\frac{\Gamma(\rightarrow\bar{\nu}_e e^- W^{*+}\rightarrow \nu_e \bar{\nu}_e e^+ e^-)}
{\Gamma(\rightarrow\nu_e \bar{\nu}_e)}
 = \frac{g^4}{3^2\,2^9\,(2\pi)^4}\,x_W^2\\\nonumber
&&\times\int_{0}^{1}dy\,
\frac{1 + 17y^3 -9y^2-9y -(6y^3 + 18y^2)\ln(y)}
{\left(y\,x_W-1\right)^2 + \left(\frac{\Gamma_W}{m_{W}}\right)^2},
\end{eqnarray}
expressed in terms of dimensionless scaling variables $y \equiv Q^2/s$ 
and $x_W\equiv s/m_W^2$.

\begin{figure}[ht]
\centering
\includegraphics[angle=270,width=1.0\columnwidth]{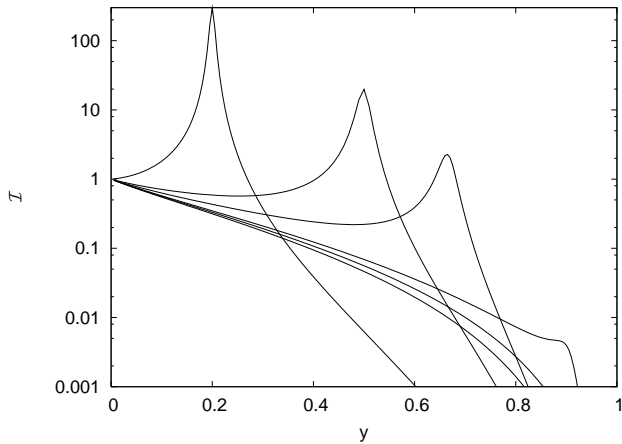}
\caption{
Integrand of Eq.~\ref{hardsingle} versus $y=Q^2/s$; in ascending order of the curves, the values
of $x_W=s/m_W^2$ are 0.9, 1.0, 1.1, 1.5, 2.0, and 5.0.
\label{fig:integrand}}
\end{figure}

In Fig.~\ref{fig:integrand} we plot the integrand of
Eq.~(\ref{hardsingle}) versus $y=Q^2/s$, for the values $x_W\equiv s/m_W^2=$
0.9, 1.0. 1.1, 1.5, 2.0, and 5.0.
The figure reveals that, for $x_W\agt 1.5$, i.e, for
$m_{\chi}\sim\sqrt{s}/2 \agt 0.6 m_W$, the rate $\Gamma(\rightarrow
\nu_e \bar{\nu}_e e^+ e^-)$ is dominated by the on-shell
$W$-resonance.  (Apparently, the $q^{-4}$ pole in Eq.~(\ref{eq:1}) is
effectively negated by the vanishing of massless three-body phase
space for $\nu\:e^+ e^-$ at $q^2=0$.)  Thus, we are justified in using
the narrow width approximation (NWA) for the $W$-propagator
\begin{eqnarray}
\frac{1}{(Q^2 - M^2) + (M\Gamma)^2}\rightarrow \frac{\pi}{M\Gamma}\ \delta(Q^2-M^2).
\end{eqnarray}
With this approximation, the cross section factorizes into the
on-shell production and subsequent decay of the $W$ boson, and the
contribution from virtual (off-shell) $W$ bosons is neglected.  Using
the NWA approximation, the integral in Eq.~(\ref{hardsingle}) is
easily evaluated.  The resulting ratio of widths becomes 
\begin{eqnarray}\label{answer1} 
&&
R_W\equiv\frac{\Gamma(\rightarrow\stackrel{{\scriptscriptstyle (-)}}{\nu}
l^\pm W^\mp\rightarrow \textrm{All})}
{\Gamma(\rightarrow \nu \bar{\nu})} \nonumber \\
&&= (2\times 9) \times 
\frac{g^4}{3^2\,2^{10}\,(2\pi)^3}\frac{m_W}{\Gamma_W} x_W 
\\ \nonumber
&&\times \left[ 1+\frac{17}{x_W^3}-\frac{9}{x_W^2}-\frac{9}{x_W} 
+\left(\frac{6}{x_W^3} + \frac{18}{x_W^2} \right)\ln(x_W)\right]\,.
\end{eqnarray}
Here we have 
dropped the flavor subscript on the (anti)neutrinos since this ratio remains
the same when flavors are summed,
and multiplied by a prefactor of $2\times 9$, which we now explain.  
The ``2'' comes from
adding the $W^-$-strahlung channel to the $W^+$ channel.  Note that
the two amplitudes do not interfere since the charges of the produced
$W$'s, and therefore of the pairs they produce with invariant mass
$M\sim m_W$, are distinguishable.  The ``9'' comes from summing over
all decay channels available to the decaying $W$.  We have three
leptonic channels, and two quark flavor channels, the latter
multiplied by three for color channels.

We may also evaluate the W-width.  At the level of our calculation, we
have for this quantity
\beq
\Gamma_W=(9)\times \frac{g^2}{48\pi}\,m_W\,.
\eeq
The ``9'' here is the same final state count that appeared in
Eq.~(\ref{answer1}).  Inputting this width into Eq.~(\ref{answer1})
and using $g^2=4\pi\alpha/\sin^2\theta_w$, we arrive at our
final expression,
\begin{eqnarray}
\label{nwa}
R_W &=& \left(\frac{\alpha}{4\pi\sin^2\theta_w}\right) \left(\frac{x_W}{48}\right) 
\\\nonumber 
&\times& \left[ 1+\frac{17}{x_W^3}-\frac{9}{x_W^2}-\frac{9}{x_W} 
   +\left(\frac{6}{x_W^3} + \frac{18}{x_W^2} \right)\ln(x_W)\right].
\end{eqnarray}
We note that this expression may also be obtained by directly computing
the production of real (on-shell) gauge bosons.  However, in that
case one must choose unitary gauge, where all degrees of freedom
are physical, in order to reproduce Eq.~(\ref{nwa}).

For our numerical work, we will take $\sin^2\theta_w=0.231$
($\sin^{-2}\theta_w=4.33$), and $\alpha=1/128$ as appropriate for
physics at the electroweak scale.  This latter choice is especially
appropriate in light of the accuracy of the NWA; $q^2$ of the virtual
neutrino will itself have a value near the threshold for on-shell $W$
production, i.e., at $\sim m_W^2$.

\section{Z-strahlung}

The cross section for the Z-strahlung process, $\chi\chi \rightarrow
\bar\nu \nu Z^* \rightarrow \textrm{All}$, may be calculated similarly
to that for W-strahlung.  For a scalar coupling as assumed in
Fig.~\ref{fig:wbrem}, there is no interference between diagrams in
which the $Z$ is radiated by the $\nu$ and $\bar\nu$.  
Thus the cross section for the Z channel is simply obtained from that
for the W channel given in Eq.~(\ref{nwa}) by dividing by a factor of
$2\cos^2\theta_W \sim 1.54$, and replacing $x_W$ with $x_Z\equiv
s/m_Z^2$, viz.
\begin{equation}
R_Z(x_Z) = \frac{1}{2\cos^2\theta_W}R_W(x_Z)\,.
\end{equation}

\section{Discussion}

\begin{figure}[th]
\includegraphics[angle=270,width=1.0\columnwidth]{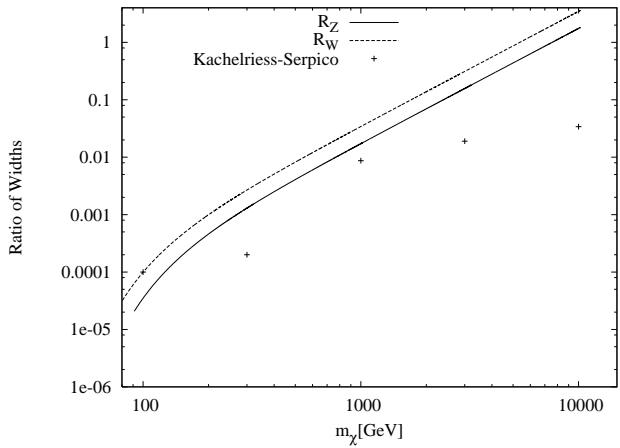}
\caption{The ratios of rates for $W$- and $Z$-strahlung to $\nu\nubar$
  production are plotted versus the dark matter mass $m_\chi$.
  Evident is the the dominance of the leading linear term in
  Eq.~\ref{answer1} above $x_G\sim 10$.  Extrapolations beyond
  $m_\chi\sim$~TeV hold some uncertainty due to multiple
  bremsstrahlung, to possible yet unknown new physics, and eventually
  to re-summation of infrared near-singularities.  Shown for
  comparison are the $R_Z$ points from~\cite{Kachelriess}.
\label{fig:final}}
\end{figure}

Figure~\ref{fig:final} shows the ratios $R_W$ and $R_Z$ as functions
of $m_\chi$.  We choose to plot rates versus $m_\chi$ rather than
$x_W$ and $x_Z$ to make the presentation more physical, and to better
illustrate the difference between the $W$ and $Z$ rates.  To convert
from the scaling variables to $m_\chi$, we have used the expressions
$x_G=s/m^2_G\approx 4\,(m_{\chi}^2/m_G^2)$, $G=W,Z$, appropriate for
non-relativistic dark matter.  The curve for $R_Z$ may be directly
compared to the points from~\cite{Kachelriess}, which we show.
Qualitative but not quantitative agreement is evident.

Let us discuss some general features of $R_G$, $G=W,Z$.  At large
$s\gg m_G^2$, we expect a leading term linear in the dimensionless
variable $x_G=s/m_G^2$.  The factor of $s$ arises from the ratio of
3-body to 2-body phase space, while the numerator is provided by the
only other dimensionful quantity in the process.  From
Fig.~\ref{fig:final}, we see that the leading linear term indeed
dominates above $x_G\sim 10$, which corresponds to $m_\chi\agt
1.5\,m_G$.  Thus, we may write a very simple expression for the width
ratio at $x_G\agt 10$.  It is
\beq
\label{simple}
R_W=
\left(\frac{\alpha}{4\pi\sin^2\theta_w}\right)\,\left(\frac{x_W}{48}\right)\,,
\quad
{\rm for}\ x_W\agt 10\,,
\eeq
and likewise times $(2\,\cos^2\theta_W)^{-1}$ for $R_Z$.  It is
unsurprising that the inequality $s\agt 10\,m^2_G \Leftrightarrow
m_{\chi}\agt 1.5\,m_G$ has appeared twice, earlier to put the $W$ or
$Z$ on-shell, and here to impose the dominance of the leading term in
the expression for $W$- or $Z$-strahlung process.

In the very large $s$ (or equivalently, the very large $m_\chi$) limit, 
the branching ratio for multi W/Z production will become sizeable.  
We may estimate the onset of double-$W/Z$ production.
The general formula for $n$-body massless phase space is 
\beq
LIPS^{(n)}= \frac{1}{8\pi}\,\left(\frac{s}{16\pi^2}\right)^{n-2}\,\frac{1}{(n-1)!\,(n-2)!}\,.
\eeq
Neglecting combinatoric factors, the perturbative expansion parameter
for additional $W/Z$~bosons is then
$(\frac{g}{\sqrt{2}})^2\,\frac{x_G}{16\pi^2}\sim
\alpha\,m_\chi^2/m_G^2$.  Thus, perturbation theory becomes unreliable
and multiple $W/Z$~production occurs when $x_G\sim 2\pi/\alpha$, or
$m_\chi\sim \alpha^{-1/2}\,m_G\sim$~TeV.  Resummations in the very
large $s$ regime, involving ordered $\ln^2 (x_G)$ terms from emission
of (nearly) massless or collinear $W/Z$'s, are discussed
in~\cite{Berezinsky}.  Co-emission of a hard photon will also occur,
at a rate comparable to double $W/Z$ emission~\cite{unitarity}.

Finally, we note that our results are easily applied to the DM decay
process $\chi\rightarrow \nu\nubar$, where $\chi$ is now a boson, with
$x_G=s/m_G^2 \rightarrow m_\chi^2/m_G^2$.  Indeed, similar expressions
will hold for any neutrino production mechanism in which the invariant
mass of the $\bar\nu \nu$ exceeds $m_W.$

\section{Conclusions}

The decay of $W$ and $Z$ bosons produced via electroweak
bremsstrahlung will lead to neutral pions and thus photons.  One may
constrain this DM annihilation signal by considering its contribution
to the Galactic or extragalactic diffuse gamma ray background.  This
was considered in Ref.~\cite{Kachelriess}, where it was shown that the
contribution of the process $\sigma(\chi\chi\rightarrow \bar\nu \nu
Z)$ to the galactic gamma ray background imposed limits on the lowest
order process, $\sigma(\chi\chi\rightarrow \bar\nu \nu)$, comparable
to those obtained directly with neutrinos.  The branching ratio
expressions we have derived differ quantitatively, through not
qualitatively, from the cross-section estimates in
Ref.~\cite{Kachelriess}.  Thus, our results lead to similar bounds.
Future Galactic gamma ray observations, such as those to be made by
GLAST, have the potential to somewhat reduce the diffuse backgrounds
through better point source identification, and to measure the
background more precisely.  In turn, this will strengthen the
electromagnetic constraint on DM annihilation, and increase the
utility of the quantitative results we have presented herein.

\section*{Acknowledgements}
We thank John Beacom, Michael Kachelriess, Pasquale Serpico, Michael
Ramsey-Musolf and Ray Volkas for helpful discussions.  NFB was
supported by the University of Melbourne Early Career Researcher and
Melbourne Research Grant Schemes, TDJ was supported by the
Commonwealth of Australia, TJW and JBD were supported in part by
U.S.~DoE grant DE--FG05--85ER40226, and TJW benefited from the
gracious hospitality of the University of Melbourne.

\section{Appendix: Initial ($\chi\chi$) or final ($\nu\nu$) state Majorana fermions}
\label{sec:appendix}
Although our goal in this paper has been to present radiative
corrections to the no-neutrino tree-level process in a manner as
model-independent as possible, it is nevertheless interesting to ask
what constraints would arise if the particles of either the intial
${\bar\chi}\chi$ state or the final $\nubar\nu$ state (or both) are
Majorana particles.  It is quite possible that neutrinos are Majorana
particles.  It is also possible that the DM is Majorana.  Although the
LSP in supersymmetric extensions of the SM is not a neutrinos-only DM
candidate, it provides a popular example of Majorana DM.  SUSY
examples of Majorana fermions include the neutralino and the photino.
(On the other hand, Kaluza-Klein DM, with mass reflecting the length
scale of extra dimensions, is a popular example of non-Majorana DM.
Typically the LKP is the bosonic recurrence of the photon.  For a
no-neutrinos model, the DM would be different again, and currently
unknown.)

Two identical fermions comprise a Majorana pair.  A fermion pair can
have total spin $S$ in the symmetric state $S=1$ or in the
antisymmetric state $S=0$.  The parity of the two-fermion state is
$P=(-)^{L+1}$, where $L$ is the orbital angular momentum of the pair.
This parity formula holds for both Dirac and Majorana pairs.  The
negative intrinsic parity of the pair, independent of the orbital
parity $(-)^L$, is the same for Dirac and Majorana pairs for different
reasons.  In the Dirac case, the $u$ and $v$ spinors (equivalently,
the positive and negative energy states) are independent and have
opposite parity corresponding to the $\pm 1$ eigenvalues of the parity
operator $\gamma^0$.  Reinterpretating the two spinor types, or
positive and negative energy states, as particle and antiparticle,
then leads directly to opposite intrinsic parity for the
particle-antiparticle pair.  In the Majorana case, the fermion has
intrinsic parity $\pm i$, and so the two-particle state has intrinsic
parity $(\pm i)^2=-1$.

On general grounds, the $L^{\rm th}$~partial wave contribution to the
annihilation rate is suppressed as $v^{2L}$, where $v$ is the relative
velocity between the heavy, non-relativistic $\chi\chi$ pair.  The
virial velocity in our Galactic halo is only $v\sim 300\ {\rm
  km/s}\sim 10^{-3}c$, so even for $L=1$ the supression is
considerable.  Thus only the $L=0$ partial wave gives an unsuppressed
annihilation rate in today's Universe.  The $L\ge 2$ states are too
suppressed to contribute to observable rates.

A Majorana pair is even under charge-conjugation
(particle-antiparticle exchange), and so from the general relation
$C=(-)^{L+S}=+1$ one infers that $L$ and $S$ must be either both even,
or both odd for the pair.  The origin of the $C=(-)^{L+S}=+1$ rule is
as follows: Under particle-antiparticle exchange, the spatial wave
function contributes $(-)^L$, and the spin wave function contributes
(+1) if in the symmetric triplet $S=1$ state, and ($-1$) if in the
antisymmetric $S=0$ singlet state, i.e., $(-)^{S+1}$.  In addition,
there is an overall ($-1$) from anticommutation of the two
particle-creation operators $b^\dag d^\dag$ for the Dirac case, and
$b^\dag b^\dag$ for the Majorana case.

Consider the $L \le 2$ states.  In spectroscopic notation
$^{(2S+1)}L_J$ and spin-parity notation ($J^{PC}$), the vector $^3
S_1$~$(1^{--}$), $C$-odd axial vector $^1 P_1$~$(1^{+-})$, and
assorted $^3 D_J$~$(J^{--})$ states are all $C$-odd and therefore
disallowed.  The pseudoscalar $^1 S_0$~($0^{-+})$, scalar $^3
P_0$~($0^{++})$, axial vector $^3 P_1$~($1^{++})$, $C$-even tensor $^3
P_2$~($2^{++})$, and pseudotensor $^1 D_2$~($2^{-+}$) are all $C$-even
and therefore allowed.  In particular, the sole $L=0$~state, with no
$v^{2L}$ suppression, is the pseudoscalar $^1 S_0$~($0^{-+})$.

Incidentally, at threshold, defined by $s=(2M_\chi)^2$ or
$v=\sqrt{1-4M_\chi ^2/s}=0$, the orbital angular momentum $L$ is
necessarily zero.  With two identical Majorana fermions, the
two-particle wave function must be antisymmetric under particle
interchange.  Since $L=0$ at threshold, the $\chi\chi$ spatial wave
function is even, and the wave function must be antisymmetrized in its
spin.  The antisymmetric spin wave function is the $S=0$ state.  Thus,
the only contributing partial wave at threshold is the $^1 S_0$ state.
We have just seen that this is also the only state with no $v^{2L}$
suppression, so one may expect an unsuppressed Majorana annihilation
rate at threshold if and only if there is a $^1 S_0$ partial wave.

One may also invoke $CP$ invariance to note that the spin $S$ of
initial and final two-fermion states, Dirac or Majorana, are the same.
This follows simply from $CP=(-)^{L+S} (-)^{L+1}=(-)^{S+1}$, and the
fact that $S=0,\,1$ are the only possibilities for a pair of spin 1/2
particles.

What does this all mean for a model with an $s$-channel exchange
particle coupling to Majorana bilinears?  It means that among the
basis fermion bilinears, the candidates are just the pseudoscalar
$\Psibar i\gamma_5\Psi$~($0^{-+})$, the scalar
$\Psibar\Psi$~($0^{++}$), and the axial vector $\Psibar\gamma^\mu
\gamma_5 \Psi$~($1^{++}$).  The vector $\Psibar\gamma^\mu
\Psi$~($1^{--}$), tensor $\Psibar\sigma^{\mu\nu}\Psi$~($2^{+-}$), and
pseudotensor $\Psibar i\gamma_5\sigma^{\mu\nu}\Psi$~($2^{--}$)
bilinears are $C$-odd and therefore disallowed.  The only $s$-channel
particles which may couple to these candidate bilinears are the
pseudoscalar, scalar, or axial vector.

There is some subtlety associated with the s-channel exchange of an
axial-vector.  The axial-vector is an $L=1$ mode, and we have
seen that this mode elicits a $v^2$ suppression in the rate.  However,
the exchange particle is off-shell (away form resonance) and so has a
timelike pseudoscalar piece in addition to the axial three-vector
piece.  This pseudoscalar coupling is effectively
$\partial_\mu\,(\Psibar\gamma^\mu\gamma_5\Psi)$.  The weak interaction
coupling of the pion to the axial vector current provides a familiar
example of such a coupling.  The axial current is not conserved, and
so the pseudoscalar coupling is nonzero.  One has
$\partial_\mu\,(\Psibar\gamma^\mu\gamma_5\Psi)=2im_\nu\Psibar\gamma_5
\Psi -\frac{\alpha_W}{\pi}\epsilon^{\mu\nu\alpha\beta}k_\mu\lambda_\nu
(k) \kbar_\alpha\lambda_\beta (\kbar)$.  The first term shows an
$m_\nu$-dependence in the amplitude, leading to $(m_\nu/M_\chi)^2$
helicity-suppression of the $L=0$ piece, while the second term is the
famous anomalous VVA coupling.  It offers $W^+W^-$ and $ZZ$ production
(with momenta $k$, $\kbar$ and helicities $\lambda(k)$,
$\lambda(\kbar)$), but at higher order $\alpha_W= g_V^2/4\pi$ in the
electroweak $W\nu\nu$ or $Z\nu\nu$ coupling $g_V$.  The linear
combination of a $v^2$-suppressed $L=1$ piece and a $m_f^2$-suppressed
$L=0$ piece to the rate from axial vector exchange was first noticed
by Goldberg~\cite{Haim1983}.

For $s$-channel exchange of a true pseudoscalar or scalar particle, 
there is no helicity suppression.
%
%
In addition, it was shown several years ago~\cite{BergstromFolk} that
helicity suppression in the axial-vector case may be avoided when the
two-body final state is replaced by a three-body final state.  In the
work of~\cite{BergstromFolk}, a charged pair was produced and a photon
was radiated from one of the charged particles.  Radiation of the
photon changes the Dirac structure of the current; it also allows one
fermion to become virtual, with a large $k^2$ replacing its small
mass.  It was found that brehmsstrahlung from the final state was
dominated by collinear and infrared emission, which left the emitting
fermion nearly on-shell, and helicity suppressed.  However, photon
emission from the accompanying $t$-channel particle did give an
unsuppressed amplitude.  In our work, it is a massive $W$ or $Z$-boson
that is radiated.  There are no collinear of infrared singularites,
and the virtual fermion is necessarily off-shell by $k^2\sim M_W^2$.
It seems likely to us that this $2\rarr 3$~rate for the radiatively
corrected $s$-channel axial-vector exchange will exceed the
helicity-suppressed $2\rarr 2$~rate by
$\frac{\alpha}{4\pi}(\frac{M_W}{m_\nu})^2$, which is many orders of
magnitude.  We do not pursue this feature of the axial-vector exchange
further here, for the premise of the neutrino-only model, investigated
in this work, is that the tree-level annihilation rate to neutrinos is
unsuppressed.

So far we have discussed $s$-channel exchange processes.  We turn now
to a brief discussion of $t$-channel exchange annihilation models.
The implications of a Majorana $\chi\chi$~pair are best recognized by
Fierz transforming the two fermion bilinears to ``charge-retention''
order, i.e., to a $\chi$-bilinear and a $\nu$-bilinear.  If the
Fierz'd bilinears contain a pseudoscalar, there is no suppression of
the rate.  Otherwise, there is a $v^2$ rate suppression.  If the
Fierz'd bilinears contain an axial vector piece, it contributes to the
rate a $v^2$-suppresed piece, and a $(m_\nu/M_\chi)^2$
helicity-suppressed piece.  (However, as just explained above, a $W$
or $Z$ radiated in the final state may well lead to a much enhanced
rate.)

It is illuminating to explain in this context the often seen remark
that LSP-annihilation has a helicity-suppressed rate to fermions.
This is true for SUSY extensions of the SM. It is not true in general
for models of dark matter.  For SUSY extensions of the SM, the
annihilation graphs consist of $t$-channel scalar exchanges, and from
crossing the identical Majorana fermions, also $u$-channel scalar
exchanges; in addition, there are scalar, pseudoscalar and
axial-vector $s$-channel exchanges.  Fierzing the $t$- and $u$-channel
scalar exchanges yields $s$-channel axial-vector
bilinears~\cite{AVfierz}, with the concomitant helicity-suppressed
$L=0$ contribution and $v^2$-suppressed $L=1$ contribution to the
annihilation rate.  The only contributions that are potentially large
come from the $s$-channel pseudoscalars.  However, the scalars and
pseudoscalars are Higgs particles, whose Yukawa couplings $g_Y$ to the
SM fermion are all proportional to $(m_f/{\rm vev})$, thereby giving
the same effect as a true helicity suppression.  Recall that the
Higgses are assigned the double burden of providing mass to the
electroweak gauge bosons and to the fermions.  The $g_Y\propto {\rm
  vev}^{-1}$ relation can be traced back to mass generation of the
gauge bosons, while the $g_Y\propto m_f$ relation comes from mass
generation for the fermions.  There are other possibilities for the
Yukawa coupling to neutrinos.  To give one example, if the neutrino
mass is small as a result of a see-saw mechanism, then the coupling of
the Higg doublet to the neutrino field will be $\sim {\rm GeV}/{\rm
  vev}$, similar to the coupling of the Higgs to most quarks.  And
importantly, a more general scalar or pseudoscalar field, not
complicit in fermon mass generation, would couple with an arbitrary
$g_Y$.

The upshot of all this for our investigation is the following: With an
$s$-channel scalar or pseudoscalar $B$-meson exchange, there is no
helicity suppression.  The Yukawa coupling of the $B$-meson to
$\nubar\nu$ (and to $\chi\chi$) is arbitrary.  The scalar exchange
proceeds in the $L=1$ partial wave, which suppresses the
$\chi\chi$~annihilation rate by $v^2$.  On the other hand, the
pseudscalar exchange proceeds in the $L=0$ partial wave, with no $v^2$
suppression of the rate.  These deductions from partial wave analysis
hold true for Dirac or Majorana $\chi$.

In the rate ratio that we investigate, any $v^2$ rate suppression
factors out.  Moreover, when our assumed scalar exchange is {\sl
  replaced with a pseudoscalar exchange, there is no change in the
  rate ratio}.  This is easily seen by noting that placement of an
$i\gamma_5$ in the amplitude of Eq.~(7) just before the spinor
$v(k_1)$ alters this amplitude by just the overall phase $-i$
(neglecting the neutrino mass).  Thus, our results as presented hold
also for the unsuppressed $L=0$ $s$-channel exchange, all the way down
in energy to the $\chi\chi$ annihilation threshold.


\begin{thebibliography}{99}
\bibitem{unitarity}
While unitarity of the DM annihilation cross-section does not directly impact
the results in this paper, we take this opportunity to summarize the unitarity 
constraints.
%
The velocity-averaged annihilation rate must satisfy 
$\langle v_{rel}\,\sigma_{ann}\rangle \le \frac{4\pi}{m_\chi^2\,v_{rel}}$.
In terms of the spinless $B$ exchange model assumed in this work,
the unitarity constraint becomes (for $m_B^2\gg m_\chi^2$)\\
$\left(\frac{M_\chi}{M_B}\right)^2
   \left(\frac{\Gamma(B\rightarrow\chi\chi)}{M_B}\right)
   \left(\frac{\Gamma(B\rightarrow{\rm SM})}{M_B}\right) \le 1$,
where SM refers to Standard Model final states.
There is no unitarity constraint on a decay width, a remark relevant 
for the case where DM decay dominates DM annihilation.
\end{thebibliography}
\end{document}